\documentstyle[12pt,epsf]{article}

\begin{document}
\makeatletter
 \renewcommand{\theequation}{%
    \thesection.\arabic{equation}}
 \@addtoreset{equation}{section}
\makeatother




\def\rmd{{\rm d}}
\def\rmD{{\rm D}}
\def\rme{{\rm e}}
\def\rmO{{\rm O}}
\def\da{\dagger}


\def\tq{\tilde{q}}
\def\bfp{{\bf p}}
\def\bfq{{\bf q}}
\def\bfr{{\bf r}}
\def\bfs{{\bf s}}
\def\bft{{\bf t}}
\def\bfu{{\bf u}}
\def\bfv{{\bf v}}
\def\bfw{{\bf w}}
\def\bfx{{\bf x}}
\def\bfy{{\bf y}}
\def\bfz{{\bf z}}


\def\rz{\blackboardrrm}
\def\gz{\blackboardzrm}
\def\Im{{\rm Im}\,}
\def\Re{{\rm Re}\,}


\def\defeq{\mathrel{\mathop=^{\rm def}}}
\def\proof{\noindent{\sl Proof:}\kern0.6em}
\def\endproof{\hskip0.6em plus0.1em minus0.1em
\setbox0=\null\ht0=5.4pt\dp0=1pt\wd0=5.3pt
\vbox{\hrule height0.8pt
\hbox{\vrule width0.8pt\box0\vrule width0.8pt}
\hrule height0.8pt}}
\def\frac#1#2{\hbox{$#1\over#2$}}
\def\dual{\mathstrut^*\kern-0.1em}
\def\mod{\;\hbox{\rm mod}\;}
\def\ring{\mathaccent"7017}
\def\lvec#1{\setbox0=\hbox{$#1$}
    \setbox1=\hbox{$\scriptstyle\leftarrow$}
    #1\kern-\wd0\smash{
    \raise\ht0\hbox{$\raise1pt\hbox{$\scriptstyle\leftarrow$}$}}
    \kern-\wd1\kern\wd0}
\def\rvec#1{\setbox0=\hbox{$#1$}
    \setbox1=\hbox{$\scriptstyle\rightarrow$}
    #1\kern-\wd0\smash{
    \raise\ht0\hbox{$\raise1pt\hbox{$\scriptstyle\rightarrow$}$}}
    \kern-\wd1\kern\wd0}


\def\nab#1{{\nabla_{#1}}}
\def\nabstar#1{\nabla\kern-0.5pt\smash{\raise 4.5pt\hbox{$\ast$}}
               \kern-4.5pt_{#1}}
\def\drv#1{{\partial_{#1}}}
\def\drvstar#1{\partial\kern-0.5pt\smash{\raise 4.5pt\hbox{$\ast$}}
               \kern-5.0pt_{#1}}


\def\momp#1#2{
    \setbox0=\hbox{${#1}$}\setbox1=\hbox{${#1}_{#2}$}
    {#1}_{#2}\kern-\wd1\kern\wd0
    \smash{\raise4.5pt\hbox{$\scriptscriptstyle +$}}}
\def\momm#1#2{
    \setbox0=\hbox{${#1}$}\setbox1=\hbox{${#1}_{#2}$}
    {#1}_{#2}\kern-\wd1\kern\wd0
    \smash{\raise4.5pt\hbox{$\scriptscriptstyle -$}}}
\def\mompm#1#2{
    \setbox0=\hbox{${#1}$}\setbox1=\hbox{${#1}_{#2}$}
    {#1}_{#2}\kern-\wd1\kern\wd0
    \smash{\raise4.5pt\hbox{$\scriptscriptstyle \pm$}}}
\def\smomp#1#2{
    \setbox0=\hbox{${#1}$}\setbox1=\hbox{${#1}_{#2}$}
    {#1}_{#2}\kern-\wd1\kern\wd0
    \smash{\raise3pt\hbox{$\scriptscriptstyle +$}}}
\def\smomm#1#2{
    \setbox0=\hbox{${#1}$}\setbox1=\hbox{${#1}_{#2}$}
    {#1}_{#2}\kern-\wd1\kern\wd0
    \smash{\raise3pt\hbox{$\scriptscriptstyle -$}}}
\def\smompm#1#2{
    \setbox0=\hbox{${#1}$}\setbox1=\hbox{${#1}_{#2}$}
    {#1}_{#2}\kern-\wd1\kern\wd0
    \smash{\raise3pt\hbox{$\scriptscriptstyle \pm$}}}
\def\mp{m_{\rm p}}
\def\si{\kern1pt{\rm si}}
\def\co{\kern1pt{\rm co}}


\def\MeV{{\rm MeV}}
\def\GeV{{\rm GeV}}
\def\TeV{{\rm TeV}}
\def\fm{{\rm fm}}

\def\fpi{F_{\pi}}


\def\euler{\gamma_{\rm E}}


\def\Nf{N_{\rm f}}
\def\psibar{\overline{\psi}}
\def\psiclass{\psi_{\rm cl}}
\def\psibarclass{\psibar_{\rm cl}}
\def\psitilde{\widetilde{\psi}}
\def\rhoprime{\rho\kern1pt'}
\def\rhobar{\bar{\rho}}
\def\rhobarprime{\rhobar\kern1pt'}
\def\rhobartilde{\kern2pt\tilde{\kern-2pt\rhobar}}
\def\rhobartildeprime{\kern2pt\tilde{\kern-2pt\rhobar}\kern1pt'}
\def\etabar{\bar{\eta}}
\def\chibar{\overline{\chi}}
\def\phibar{\overline{\phi}}
\def\zetabar{\bar{\zeta}}
\def\zetaprime{\zeta\kern1pt'}
\def\zetabarprime{\zetabar\kern1pt'}
\def\zetar{\zeta_{\raise-1pt\hbox{\sixrm R}}}
\def\zetabarr{\zetabar_{\raise-1pt\hbox{\sixrm R}}}
\def\phieff{\phi_{\rm eff}}
\def\phiimpr{\phi_{\kern0.5pt\hbox{\sixrm I}}}
\def\phir{\phi_{\hbox{\sixrm R}}}
\def\ar{A_{\hbox{\sixrm R}}}
\def\pr{P_{\hbox{\sixrm R}}}
\def\sr{S_{\hbox{\sixrm R}}}
\def\ce{{\cal E}}


\def\dirac#1{\gamma_{#1}}
\def\diracstar#1#2{
    \setbox0=\hbox{$\gamma$}\setbox1=\hbox{$\gamma_{#1}$}
    \gamma_{#1}\kern-\wd1\kern\wd0
    \smash{\raise4.5pt\hbox{$\scriptstyle#2$}}}


\def\bx{b_{\rm X}}
\def\ba{b_{\rm A}}
\def\bp{b_{\rm P}}
\def\bv{b_{\rm V}}
\def\bs{b_{\rm S}}
\def\bt{b_{\rm T}}
\def\bap{b_{\rm A,P}}
\def\bg{b_{\rm g}}
\def\bm{b_{\rm m}}
\def\bzeta{b_{\zeta}}
\def\bmSF{b_{\rm m}^{\hbox{\sixrm SF}}}

\def\cx{c_{\rm X}}
\def\ca{c_{\rm A}}
\def\cv{c_{\rm V}}
\def\Ca{\ca}
\def\Cv{\cv}
\def\cT{c_{\rm T}}
\def\csw{c_{\rm sw}}
\def\cs{c_{\rm s}}
\def\ct{c_{\rm t}}
\def\cst{c_{\rm s,t}}
\def\ctildes{\tilde{c}_{\rm s}}
\def\ctildet{\tilde{c}_{\rm t}}
\def\ctildest{\tilde{c}_{\rm s,t}}


\def\fa{f_{\rm A}}
\def\fda{f_{\delta{\rm A}}}
\def\fp{f_{\rm P}}
\def\fx{f_{\rm X}}
\def\kv{k_{\rm V}}
\def\kt{k_{\rm T}}
\def\f1{f_1}
\def\hx{h_{\rm X}}
\def\ha{h_{\rm A}}
\def\hda{h_{\rm dA}}
\def\hp{h_{\rm P}}
\def\hv{h_{\rm V}}
\def\h1{h_1}


\def\SUtwo{{\rm SU(2)}}
\def\SUthree{{\rm SU(3)}}
\def\SUn{{\rm SU}(N)}
\def\tr{\,\hbox{tr}\,}
\def\Ad{{\rm Ad}\,}
\def\CF{C_{\rm F}}
\def\cf{\CF}


\def\Sg{S_{\rm G}}
\def\Sf{S_{\rm F}}
\def\Seff{S_{\rm eff}}
\def\Simpr{S_{\rm impr}}
\def\Simprf{S_{\rm F\!,impr}}
\def\Zf{{\cal Z}_{\rm F}}
\def\op#1{{\cal O}_{\rm #1}}
\def\opprime#1{\setbox0=\hbox{${\cal O}$}\setbox1=\hbox{${\cal O}_{\rm #1}$}
    {\cal O}_{\rm #1}\kern-\wd1\kern\wd0
    \smash{\raise4.5pt\hbox{\kern1pt$\scriptstyle\prime$}}\kern1pt}
\def\ophat#1{\widehat{\cal O}_{\rm #1}}
\def\ophatprime#1{\setbox0=\hbox{$\widehat{\cal O}$}
    \setbox1=\hbox{$\widehat{\cal O}_{\rm #1}$}
    \widehat{\cal O}_{\rm #1}\kern-\wd1\kern\wd0
    \smash{\raise4.5pt\hbox{\kern1pt$\scriptstyle\prime$}}\kern1pt}
\def\bop#1{{\cal L}_{\rm #1}}
\def\bopprime#1{\setbox0=\hbox{${\cal O}$}\setbox1=\hbox{${\cal O}_{\rm #1}$}
    {\cal L}_{\rm #1}\kern-\wd1\kern\wd0
    \smash{\raise4.5pt\hbox{\kern1pt$\scriptstyle\prime$}}\kern1pt}
\def\blag#1{{\cal B}_{#1}}
\def\blagprime#1{\setbox0=\hbox{${\cal B}$}\setbox1=\hbox{${\cal B}_{#1}$}
    {\cal B}_{#1}\kern-\wd1\kern\wd0
    \smash{\raise5.2pt\hbox{\kern1pt$\scriptstyle\prime$}}\kern1pt}


\def\alphaSF{\alpha_{\rm SF}}
\def\alphaTP{\alpha_{\rm TP}}
\def\alphaMSbar{\alpha_{\msbar}}

\def\gms{g_{\ms}}
\def\gbar{\bar{g}}
\def\gbarMS{\gbar_{\ms}}
\def\gbarMSbar{\gbar_{\msbar}}
\def\gbarSF{\gbar_{\rm SF}}
\def\gbarTP{\gbar_{\rm TP}}
\def\gr{g_{\rm R}}
\def\gR{\gr}
\def\glat{g_{\lat}}
\def\gSF{g_{{\hbox{\sixrm SF}}}}
\def\gp{g_{\rm P}}

\def\mq{m_{\rm q}}
\def\mqtilde{\widetilde{m}_{\rm q}}
\def\mr{m_{{\rm R}}}
\def\mc{m_{\rm c}}
\def\Kc{K_{\rm c}}
\def\mp{m_{\rm p}}
\def\Mp{\mp}
\def\mlat{m_{\lat}}
\def\mSF{m_{\hbox{\sixrm SF}}}

\def\zx{Z_{\rm X}}
\def\za{Z_{\rm A}}
\def\zv{Z_{\rm V}}
\def\zp{Z_{\rm P}}
\def\zg{Z_{\rm g}}
\def\zm{Z_{\rm m}}
\def\zgm{Z_{\rm g,m}}
\def\zphi{Z_{\phi}}
\def\zmSF{Z_{\rm m}^{\hbox{\sixrm SF}}}
\def\zmlat{Z_{\rm m}^{\rm lat}}
\def\zzeta{Z_{\zeta}}

\def\Zx{\zx}
\def\Za{\za}
\def\Zv{\zv}
\def\Zp{\zp}
\def\Zg{\zg}
\def\Zm{\zm}
\def\Zgm{\zgm}
\def\Zz{\zzeta}

\def\gtilde{\tilde{g}_0}
\def\mtilde{\widetilde{m}_0}

\def\ms{{\rm MS}}
\def\msbar{{\rm \overline{MS\kern-0.05em}\kern0.05em}}
\def\lat{{\rm lat}}
\def\SF{\rm SF}
\def\ab{\bar{a}}

\title{
{\vspace{-0cm} \normalsize
\hfill \parbox{38mm}{MPI-PhT/98-48}}\\[25mm]
Computation of the improvement coefficient $\csw$ to 1-loop
with improved gluon actions}
\author
{Sinya Aoki\thanks{On leave from Institute of Physics, University of Tsukuba,
Tsukuba, Ibaraki-305, Japan} ,
Roberto Frezzotti and Peter Weisz \\
Max-Planck-Institut f\"ur Physik \\
F\"ohringer Ring 6, D-80805 M\"unchen, Germany}
\date{\today}
\maketitle
\begin{abstract}
 The coefficient $\csw$ appearing in the
 Sheikholeslami-Wohlert improved action
 is computed to one loop perturbation theory
 for improved gluon actions including six-link loops.
 The O($a$) improvement coefficients for the dimension three
 isovector composite operators bilinear in the quark fields
 are also computed to one loop order of perturbation theory
 with degenerate non-vanishing quark masses. 
\end{abstract}

\vfill
\eject

\section{Introduction}

Despite the fact that the quenched approximation is a drastic
modification of the theory, it turns out that 
the low lying hadronic spectrum
obtained in this approximation is quantitatively very close
to the experimentally observed one. Only recently, 
with the progress of lattice QCD simulations, have the  
precise results from CP-PACS \cite{Yoshie, CppacsQ}
shown a systematic deviation not only in the
meson sector but also in the baryon masses. 
It is thus now clear that one has to incorporate the effect 
of dynamical quarks to have a chance to obtain 
the correct hadron spectra in the continuum limit.
On the other hand, full QCD simulations are extremely computer time 
consuming compared to those of quenched QCD. Although the fastest 
computers now available have peak speeds of a few hundred Gflops,
it is still difficult to reduce the lattice spacing less than 0.1 fm 
while keeping the physical volume larger than 2 fm with reasonable statistics.
Due to the large scaling violation of the combination of the plaquette
gluon action and the Wilson quark action 
there is no hope 
to obtain accurate and reliable hadron spectra in the continuum limit 
by the extrapolation from such lattice spacings.

One of the possible ways to overcome this difficulty is, of course, the use
of improved actions to reduce the scaling violation. Among the several 
proposals, the least ambitious but most practical one 
known at present is the on-shell O($a$) 
improvement program, where only one new term (clover term) has to be added
to the quark action (clover action) \cite{SW}.
In this program the scaling violation is reduced to O($a^2$) with
the appropriate value of the coefficient of the clover term (clover
coefficient).
Although no modification for the plaquette gluon action is necessary in the
O($a$) improvement program, one can still add one or more terms to
the plaquette gluon action without significantly increasing CPU time, 
in order to reduce non-negligible O($a^2$) 
errors present at lattice spacings equal to or larger than 0.1 fm. 
Recently such a combination (the so-called renormalization group 
(RG) improved gluon action \cite{Iwasaki} and the clover action)
has been compared to other combinations 
at a lattice spacing of $\sim 0.2$ fm and
it has been shown that this combination has some better properties than
others\cite{CppacsF}:
the rotational symmetry of the static quark potential is well restored
and the scaling violation in hadron spectra is reasonably small.

Because of this result, the CP-PACS collaboration 
decided to use the RG improved
gluon action and the clover action for their production run of full QCD 
simulations \cite{CppacsF}
whose lattice spacings range from 0.2 fm to 0.1 fm.
Unfortunately at the time they started the production runs
the value of clover coefficient $\csw$ for the RG improved gluon
action was known neither 
non-perturbatively nor at 1-loop order of perturbation theory.
Instead they used a ``perturbative mean field'' value \cite{Parisi} 
$\csw = (1-0.8412\beta^{-1} )^{-3/4}$, where $1-0.8412\beta^{-1}$ is
the one-loop value for the plaquette.
Although this value might be a dominant contribution at 1-loop order,
it is of course necessary to know $\csw$ fully at 1-loop order
to be able to estimate how large the errors in this approximation are.
To obtain O($a$) improvement for correlation functions of composite
operators one also has to construct O($a$) improved operators
\cite{paperI}.
For the case of quark bilinear operators of dimension 3 
(see subsection 2.4) this involves computing mixing coefficients
($c_{\rm X}$, ${\rm X=A,V,T}$) as well as 
mass dependent correction factors ($1+am b_{\rm Y}$, 
${\rm Y=S,P,A,V,T,m}$).
It would of course also be desirable to push O($a$) improvement to
the non-perturbative level as it has been done by the Alpha
Collaboration \cite{Rainer} for the plaquette gluon action.

In this paper we have calculated $\csw$, $c_{\rm X}$, 
and $b_{\rm Y}$ at 1-loop
order of perturbation theory for gluon actions including six-link loops
and the clover quark action. Our computational method uses the
Schr\"{o}dinger functional \cite{alphaI}. 
Our results for the coefficients
$c_{\rm X}$, and $b_{\rm Y}$ are in 
complete agreement with the results of the 
recent computation of Taniguchi and Ukawa \cite{TaUk} who used a 
completely different method.


\section{Definitions}

As mentioned in the introduction we are here concerned with 
constructing O($a$) on-shell improved actions for QCD
which are to be used in large scale numerical simulations.
Starting with the original Wilson action $S_{\rm W}$ for
the quarks, Sheikholeslami and Wohlert \cite{SW} have shown
that O($a$) on-shell improvement can be achieved by 
adding just one extra term $\propto S_{\rm SW}$ to the action. 
Thus we will be considering total actions of the form
\begin{equation}\label{action}
S[U,\bar{\psi},\psi]=S[U]+S_{\rm W}[U,\bar{\psi},\psi]+
\csw(g_0)S_{\rm SW}[U,\bar{\psi},\psi],
\end{equation}
with $\Nf\ge2$ flavors. The part of the action involving the
quarks will be specified in more detail below.
At this point we only
stress that the factor $\csw$ multiplying the SW term 
is a function
of the bare coupling $g_0^2$ which depends on the particular
form of the gauge action $S[U]$ chosen. 
The tree-level value is the same for all gauge actions
$\csw(0)=1$ \cite{SW}
\footnote{Here we take Wilson's $r$-coefficient
fixed to be 1, otherwise $\csw(0)=r$}
and so $\csw$ has a perturbative expansion of the form
\begin{equation}
\csw (g_0)=1+\csw^{(1)} g_0^2 +\dots
\end{equation}
For the Wilson plaquette action the 1-loop coefficient  
$\csw^{(1)}$ was computed a long time ago by Wohlert 
\cite{Wohlert} (and checked in ref.~\cite{paperII}), 
and recently non-perturbatively
for the quenched theory \cite{paperIII} and full QCD
with $\Nf=2$ flavors \cite{KarlRainer}.   

\begin{figure}[htb]
\begin{center}
\leavevmode
\epsfxsize=90mm
\epsfbox{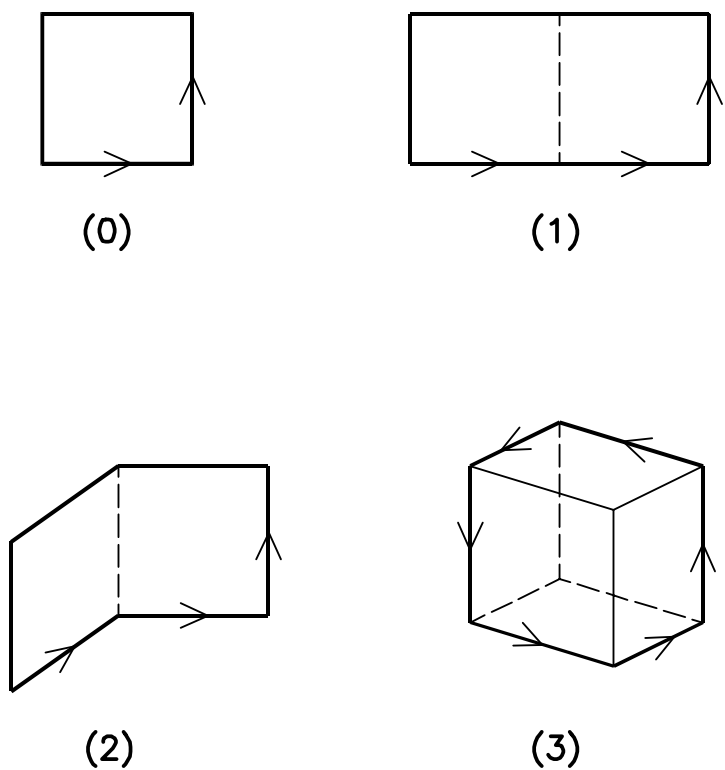}
\vskip 10mm
\end{center}
\caption{Elementary loops in ${\cal S}_i,\,\,i=0,1,2,3$}
\label{fig:1loop}  
\end{figure}

In this paper we will compute $\csw^{(1)}$ for some gauge actions 
belonging to a general class containing
loops up to length 6
\footnote{ Our notation differs from
refs.~\cite{LueWeI},\cite{LueWeII}
by an interchange of $c_2$ with $c_3$}
\begin{equation}
S[U]={2\over g_0^2}\sum_{i=0}^3 c_i(g_0^2)  
\sum_{{\cal C}\in {\cal S}_i} {\cal L}({\cal C})
\end{equation}
where the ${\cal S}_i$ denote sets of elementary loops ${\cal C}$
on the lattice as given in fig.~\ref{fig:1loop}.
Loops ${\cal C}$ that differ by orientation only are
considered equal. Furthermore
\begin{equation}
{\cal L}({\cal C})={\rm Re Tr}[I-U({\cal C})],
\end{equation}
$U({\cal C})$ being the ordered product of the link
variables $U_{\mu}(x)$ along ${\cal C}$. The coefficients
$c_i(g_0^2)$ are regular at $g_0^2=0$ and are normalized
such that
\begin{equation}
c_0(g_0^2)+8c_1(g_0^2)+16c_2(g_0^2)+8c_3(g_0^2)=1.
\end{equation}

This class of gauge actions was first considered by Wilson
\cite{WilsonI} in the framework of block spin
renormalization and includes in particular 
the action proposed by Iwasaki \cite{Iwasaki}. 
Among this class are also the on-shell 
O($a^2$) improved pure gauge actions \cite{LueWeI}.
However we stress that here we do not have the ambition to achieve
on-shell O($a^2$) improvement for full QCD. The latter
ambitious program would
involve the addition of various other terms to the action
including in particular 4-fermion operators (of dimension 6)
which would be very awkward for numerical simulations.

In our explicit computations we have only considered actions
with $c_2=0$. Apart from the Wilson action ($c_1=c_3=0$), we have 
studied 4 actions
\begin{eqnarray}
{\rm LW}: c_1(0)=&-1/12,\,\,\, c_3(0)=0 \qquad \protect{\cite{LueWeI}} ,
&\\
\noalign{\vskip2ex}
{\rm RG1}: c_1=&-0.331,\,\,\, c_3=0 \qquad \protect{\cite{Iwasaki}},
\qquad 
&\\
\noalign{\vskip2ex}
{\rm RG2}: c_1=&-0.27,\,\,\, c_3=-0.04 \qquad \protect{\cite{Iwasaki}} ,
&\\
\noalign{\vskip2ex}
{\rm RG3}: c_1=&-0.252,\,\,\, c_3=-0.17\qquad \protect{\cite{WilsonI}} .
\end{eqnarray}
Note for the LW action the $c_i$ are functions
of the bare coupling, whereas for the RG actions the $c_i$ are
constants.
 
In ref.~\cite{LueWeII}
it was shown that  demanding positivity of the action in the
limit $g_0\to 0$ restricts the
tree level coefficients
to some convex domain $P$. A sufficient condition 
for the coefficients to lie in $P$ is
\begin{equation}
c_0(0)+8\hat{c}_1+32\hat{c}_2+{80\over 3}\hat{c}_3 >  0
\end{equation}
where
\begin{equation}
\hat{c}_i=\frac12 \Bigl(c_i(0)-|c_i(0)|\Bigl). 
\end{equation}
Note that all the actions listed above satisfy this condition
except RG3.

\subsection{The Schr\"{o}dinger Functional}

The work of the Alpha Collaboration has demonstrated that
a convenient framework in order to compute improvement coefficients
and renormalization constants (and of course more importantly
many quantities of direct physical relevance) is provided by
the Schr\"odinger functional (SF) \cite{alphaI, StefanI, paperI}.
Here one considers the theory defined on 
hypercubic lattices of volume $L^3\times T$ 
with cylindrical geometry, i.e. periodic-type boundary
conditions in the spatial directions and Dirichlet boundary
conditions in the ``time" direction, for both gluon
and quark fields $U_{\mu},\psi$. From now on we set the
lattice spacing $a=1$ in most formulae.

For the case of the Wilson action the SF was described in
detail in refs.~\cite{alphaI, StefanI, paperI}. 
The extension of the SF considerations to gauge actions 
belonging to the class above was first considered
by Klassen \cite{Klassen}. We found it 
conceptually more convenient to set things up slightly differently. 
In particular in our construction the 
dynamical variables to be integrated over
are chosen independently of the form of the action: they are the
spatial link variables and quark fields $U_k(x),\psi(x)$ 
with times $x_0=1,..,T-1$
and the link variables $U_0(x)$
with times $x_0=0,..,T-1$ (i.e. inside the cylinder).
Dirichlet boundary conditions are imposed on the
fields $U_k(x),\psi(x)$ at the boundaries $x_0=0$ and $x_0=T$.

What remains to be done is to specify the action in more detail. 

\subsection{The pure gauge action $S[U]$}

Taking account of the fact that we  
are working with the Schr\"{o}dinger functional geometry,
we modify the gauge action above to read
\begin{equation}
S[U]={2\over g_0^2}\sum_{i=0}^3 
\sum_{{\cal C}\in {\cal S}_i}
W_i({\cal C},g_0^2){\cal L}({\cal C})
\end{equation}
with weights $W_i({\cal C},g_0^2)$ which may differ
from $c_i(g_0^2)$ for loops ${\cal C}$ near the boundaries.  
First we must specify the elements of the classes ${\cal S}_i$ more
precisely. These consist of all loops of the given shape which
can be drawn on the cylindrical lattice that involve only the
``dynamical links" in the sense specified above, and 
spatial links on the boundaries at $x_0=0$ and $x_0=T$.
In particular rectangles protruding out of the cylinder
are not included and hence we do not have to specify further
boundary conditions for link variables outside the cylinder.

The most important point now is that to achieve O($a$)
improvement for all (on-shell) Green functions, 
a careful choice of the weights $W_i$ is necessary
\footnote{For improvement of some physical quantities
arising from local Ward identities a particular choice of $W_i$ 
may not be necessary.}. 
From general considerations \cite{alphaI}
only two independent boundary O($a$) counterterms 
involving the gauge fields 
$\int \rmd^3x F_{0k}(x)F_{0k}(x)$ and
$\int \rmd^3x F_{kl}(x)F_{kl}(x)$ at $x_0=0$ and $x_0=T$
are expected. Since we now have many types of loops
at our disposal there are various
ways to chose the weights to achieve our goal. 
A rather natural way, which we call Choice A, is to adjust the 
O($a$) counterterms just through the plaquettes:
\begin{equation}\label{cha0}
W_0({\cal C},g_0^2)=\cases   
{&$\cs(g_0^2)$ if ${\cal C}$ lies completely on one of the boundaries, \cr
 &$\ct(g_0^2)$ if ${\cal C}$ just touches one of the boundaries, \cr
 &$c_0(g_0^2)$ otherwise,\cr}
\end{equation}
and for the other classes ${\cal S}_i,\,i=1,2,3$ we simply set:
\begin{equation}\label{chai}
W_i({\cal C},g_0^2)=\cases
{&$0$ if ${\cal C}$ lies completely on one of the boundaries, \cr
 &$c_i(g_0^2)$ otherwise.\cr}
\end{equation}
With this particular choice 
for complete O($a$) improvement we find to tree level
\begin{eqnarray}
\cs(0)=&\frac12,\quad\quad\qquad\label{cs}
&\\
\noalign{\vskip2ex}
\ct(0)=&c_0(0)+2c_1(0),\label{ct}
&
\end{eqnarray}
independently of the (smooth) boundary conditions.
The necessity for the extra term in $\ct(0)$ involving the
coefficient of the rectangles is explained in Appendix A.

From the discussion in Appendix A it will also be clear
that another admissible special way to set things up, which we call Choice
B, is to keep eq.(\ref{cha0}) but change eq.(\ref{chai}) to
\begin{equation}
W_i({\cal C},g_0^2)=\cases
{&$0$ if ${\cal C}$ lies completely on one of the boundaries, \cr
 &$\frac32 c_i(g_0^2)$ if $i=1$ and ${\cal C}$ has exactly 2 links
on a boundary,\cr
 &$c_i(g_0^2)$ otherwise.\cr}
\end{equation}
Then eq.(\ref{cs}) remains valid but instead 
of eq.(\ref{ct}) we must set
\begin{equation}
\ct(0)=c_0(0).
\end{equation}
In other words in this case one of the O($a$) boundary counterterms
at tree level is adjusted by the coefficient of the
rectangular loops of space-length 2 lattice units. The reason for
mentioning this choice will become manifest below. 

We must still specify the boundary conditions for the gauge field. 
As in the determination of the running coupling ~\cite{alphaI} we chose
\begin{equation}
U_k(x)|_{x_0=0}=e^C,\,\,U_k(x)|_{x_0=T}=e^{C^{\prime}},
\end{equation}
where $C,C^{\prime}$ are diagonal traceless antihermitian
matrices.
As explained in  ref.~\cite{alphaI} the presence of non-trivial
boundary conditions induces a constant chromo-electric background
field. In the continuum limit the background field is given 
(up to gauge transformations) by
\begin{eqnarray}
A_0(x)=&0,\,\,\,A_k(x)=b(x_0),\,\,k=1,2,3,
&\\
\noalign{\vskip2ex}\label{bfield}
b(x_0)=&{1\over T}\Bigl[ (T-x_0)C+x_0C^{\prime}\Bigr].\quad\quad
&   
\end{eqnarray}
On the lattice one will have a solution to the field equations
of the form
\begin{equation}
U_0(x)=1,\,\,U_k(x)=V(x_0),\,\,k=1,2,3,
\end{equation}
but for actions having $c_1\ne0$ and generic choices of (admissible)
boundary weights it is not possible to solve the lattice equations of
motion analytically and one must be content with numerical
solutions. Although this is of no relevance for numerical simulations,
for perturbative computations it is more practical to have
analytic solutions at tree level. This is where the special Choice
B comes in to play since it is easy to check that in this case
\begin{equation}\label{vfield}
V(x_0)=\exp b(x_0),
\end{equation}
with $b$ as in eq.(\ref{bfield}), is an exact solution (in the limit
$g_0=0$).
On the other hand for Choice A, although the above form of $V$
is a solution to the equations obtained by the variation of links
well inside the cylinder, it fails to be a solution for 
equations obtained by the variation of spatial links at
time 1 (or at time $T-1$). Thus for non-vanishing 
Abelian background field it is clearly advantageous to use
Choice B for the analytical computations.

The Feynman rules are now easily derived 
by expanding around the background field solution
\begin{equation}
U_0(x)=\exp [g_0q_0(x)],\,\,\,U_k(x)=V(x_0)\exp [g_0q_k(x)].
\end{equation}
For the validity of the perturbative expansion it is 
necessary to know whether the background field is the 
unique absolute minimum of the action (up to gauge transformations).
For the case of the Wilson plaquette action 
this was proven ~\cite{alphaI}
provided the diagonal elements of $C$ lie in a certain
so called fundamental domain. The extension of the proof 
of this theorem to
the general class of actions considered here has to our 
knowledge not been achieved. Klassen \cite{Klassen} has
however investigated the question numerically and found no
evidence to the contrary for a large class of actions. 
In fact the only cases where Klassen found
that the global minimum was not given 
by eqs.(\ref{vfield}) and (\ref{bfield}) were
those with coefficients not in the positivity domain $P$
alluded to above. We hope therefore that perturbation theory
remains trustworthy. As a consistency check   
in our computations we monitored the
eigenvalues of the quadratic gauge operator
(the inverse free propagator) and checked 
(for various covariant gauges) that it did 
not develop a negative eigenvalue. This held even for the
action RG3 which is not manifestly positive.

Concerning gauge fixing we used the covariant gauge fixing
specified in ref.~\cite{alphaI}
\begin{equation}\label{gaugefix}
S_{\rm gf}=\lambda_0 (d^* q,d^* q)
\end{equation}
with
\footnote{The reader is requested to consult ref.~\cite{alphaI}
for the specific notations.
A more symmetric way to fix the gauge
is discussed in the paper by Narayanan and Wolff \cite{NaWo}.}
\begin{equation}
(d^* q)_{\alpha\beta}(x)=\cases
{&$\sum_{\mu} (D^*_{\mu}q_{\mu})_{\alpha\beta}(x)$
 for $0<x_0<T$, \cr
 &$L^{-3}\sum_{\bf y}
 q_0(0,{\bf y})_{\alpha\beta}\delta_{\alpha\beta}$
 for $x_0=0$,\cr
 & 0 for $x_0=T$.\cr}
\end{equation}
The inverse gauge propagator for the case of zero
background field is given in Appendix B.
Note that the simple choice of gauge fixing 
given above has the disadvantage 
that there is for generic $c_i$ no choice of the 
gauge parameter where the propagator becomes diagonal
with respect to the Lorentz indices.
It may be possible to introduce other choices of gauge
fixing where this is the case but we did not 
investigate this possibility further.
In all cases we obtained the free propagator
numerically although analytical
expressions are probably available for the case 
of vanishing background field. 
The expressions for the vertices are rather lengthy
and we do not give them here, they can be obtained 
from the authors upon request.

\subsection{The fermion action}

As mentioned above for fermion action we consider only the on-shell
O($a$) improved Wilson action which is obtained by adding
the  Sheikholeslami-Wohlert interaction term
(with coefficient $\csw$) \cite{SW}. That is in eq.(\ref{action})
$S_{\rm W}$ is the standard Wilson fermion action with $r=1$,
\begin{equation}\label{W_action}
S_{\rm W}[U,\bar{\psi},\psi]=\sum_{x} \bar{\psi}(x)(D+m_0)\psi (x)
+S_{\rm Wb},
\end{equation}
where $m_0$ is the bare quark mass and
\begin{equation}
D=\frac12 \sum_{\mu}\{\gamma_{\mu}(\nabla_{\mu}^*+\nabla_{\mu})
-\nabla_{\mu}^*\nabla_{\mu}\},
\end{equation}
is the lattice Wilson Dirac operator. 
$\nabla_{\mu},\nabla_{\mu}^*$ are the forward and backward lattice
covariant derivatives
\begin{eqnarray}
\nabla_{\mu}\psi(x) =& \lambda_{\mu}U_{\mu}(x)\psi(x+\hat{\mu})-\psi(x),
&\\
\noalign{\vskip2ex}
\nabla_{\mu}^*\psi(x) =& \psi(x)-\lambda^*_{\mu }
U_{\mu}(x-\hat{\mu})^{\da}\psi(x-\hat{\mu}).
&
\end{eqnarray}
The additional $U(1)$ constant field
\begin{equation}
  \lambda_{\mu}=\cases{1      & if $\mu=0$,\cr
                      \noalign{\vskip1ex}
                      \rme^{i\theta/L} &if $\mu>0$, \cr}
\end{equation}
corresponds to a free phase in the spatial boundary conditions of the
fermions. Finally $S_{\rm Wb}$ in (\ref{W_action}) 
is an additional boundary term \cite{paperI}
which is needed to remove O($a$) lattice artifacts in the
SF framework:
\begin{eqnarray}
S_{\rm Wb} &=
(\ctildet-1)\sum_x\Bigl\{
\delta_{x_0,1}\psibar (x) 
\Bigl[\psi(x)-U_0(x-\hat{0})^{-1}P_+\psi(x-\hat{0})\Bigr]
&\nonumber\\
\noalign{\vskip2ex}
&+\delta_{x_0,T-1}\psibar (x) 
\Bigl[\psi(x)-U_0(x)P_-\psi(x+\hat{0})\Bigr]\Bigr\},
&
\end{eqnarray}
where $P_{\pm}=\frac12(1\pm \gamma_0)$.

$S_{\rm SW}$ is the Sheikholeslami-Wohlert \cite{SW} term: 
\begin{equation}
S_{\rm SW}[U,\bar{\psi},\psi]
=\frac{i}{4} \sum_{x,\mu,\nu} \bar{\psi}(x)
\sigma_{\mu\nu}{\cal F}_{\mu\nu}(x)\psi(x),
\end{equation}
where the field ${\cal F}_{\mu\nu}$ 
is given explicitly by  
\begin{eqnarray}
{\cal F}_{\mu\nu}(x)=\frac18 \Bigl\{[
&U_{\mu}(x)U_{\nu}(x+\hat{\mu})U_{\mu}(x+\hat{\nu})^{\da}U_{\nu}(x)^{\da}
&\nonumber\\
\noalign{\vskip2ex}
+&U_{\nu}(x)U_{\mu}(x+\hat{\nu}-\hat{\mu})^{\da}
U_{\nu}(x-\hat{\mu})^{\da}U_{\mu}(x-\hat{\mu})
&\nonumber\\
\noalign{\vskip2ex}
+&U_{\mu}(x-\hat{\mu})^{\da}U_{\nu}(x-\hat{\mu}-\hat{\nu})^{\da}
U_{\mu}(x-\hat{\nu}-\hat{\mu})U_{\nu}(x-\hat{\nu})
&\nonumber\\
\noalign{\vskip2ex}
+&U_{\nu}(x-\hat{\nu})^{\da}U_{\mu}(x-\hat{\nu})
U_{\nu}(x+\hat{\mu}-\hat{\nu})U_{\mu}(x)^{\da}]
&\nonumber\\
\noalign{\vskip2ex}
-&[\mu\leftrightarrow\nu]\Bigr\}.
&
\end{eqnarray}

The boundary conditions for the fermion fields have first
been specified in ref.~{\cite{StefanI}; in all our computations
we have set the quark field on the boundaries to zero.

A general introduction to on-shell O($a$) improvement 
of lattice QCD with Wilson quarks can be found in 
refs.~\cite{letter, nonpert}.
In particular there it has been discussed how
the renormalization procedure can be carried out in a way consistent
with O($a$) improvement. In the following we adopt 
the conventions and notations as introduced in this reference. 

\subsection{Composite operators of dimension 3}

On-shell O($a$) improvement requires also the introduction of
O($a$) counterterms for the composite fields.
Here we restrict attention
to a few gauge invariant composite operators which are bilinear
in the quark fields. Assuming $\Nf\ge 2$ mass degenerate
quark flavors we consider the local isovector
fields~\footnote{
  The Dirac gamma matrix conventions are as in
  Appendix A of ref.~\cite{paperI}.},
\begin{eqnarray}
  V^a_{\mu}(x)=&\bar{\psi}(x) \gamma_{\mu}\frac12\tau^a\psi (x),
  &\\
  \noalign{\vskip2ex}
  A^a_{\mu}(x)=&\bar{\psi}(x) \gamma_{\mu}\gamma_5\frac12\tau^a\psi (x),
  &\\
  \noalign{\vskip2ex}
  P^a (x)=&\bar{\psi}(x) \gamma_5\frac12\tau^a\psi (x),
  &\\
  \noalign{\vskip2ex}
  T^a_{\mu\nu}(x)=&i\bar{\psi}(x) \sigma_{\mu\nu}\frac12\tau^a\psi (x).
  &
\end{eqnarray}
Using these definitions and following ref.~\cite{paperI} the
corresponding improved vector, axial vector and tensor
currents may be parametrized as follows,  
\begin{eqnarray}
(V_{\rm I})^a_{\mu}=&V^a_{\mu}+\cv(g_0^2) 
a\tilde{\partial}_{\nu}T^a_{\mu\nu},\quad\quad\quad
&\\
\noalign{\vskip2ex}
(A_{\rm I})^a_{\mu}=&A^a_{\mu}+\ca(g_0^2) 
a\tilde{\partial}_{\mu}P^a,\quad\quad\quad
&\\
\noalign{\vskip2ex}
(T_{\rm I})^a_{\mu\nu}=&T^a_{\mu\nu}+\cT(g_0^2) 
a\Bigl( \tilde{\partial}_{\mu}V^a_{\nu}
-\tilde{\partial}_{\nu}V^a_{\mu}\Bigr),
&
\end{eqnarray}
where $\tilde{\partial}$ denotes the symmetric lattice derivative
$\tilde{\partial}_{\mu}f(x)=\frac12 [f(x+a\hat{\mu})-f(x-a\hat{\mu})]$.

Throughout the paper we will use a mass-independent renormalization
scheme which is compatible with O($a$) improvement.
In such a scheme the renormalized coupling and quark mass
are given by ~\cite{paperI}
\begin{eqnarray}
  \gr^2=&\tilde{g}_0^2Z_{\rm g}(\tilde{g}_0^2,a\mu),
  &\\
  \noalign{\vskip2ex}
  m_{\rm R}=&\tilde{m}_{\rm q}\zm(\tilde{g}_0^2,a\mu),
  &
\end{eqnarray}
where $\mu$ is the renormalization scale. 
The parameters $\tilde{g}_0$
and $\tilde{m}_{\rm q}$ are defined as follows
\begin{eqnarray}
  \tilde{g}_0^2=&g_0^2 \bigl[ 1+\bg(g_0^2)a\mq \bigr],
 \qquad\qquad\qquad\qquad
  &\\
  \noalign{\vskip2ex}
  \tilde{m}_{\rm q}=&\mq\bigl[ 1+\bm(g_0^2)a\mq \bigr],\qquad \mq=m_0-\mc,
  &
\end{eqnarray}
where $\mc$ is the critical bare quark mass.
The critical hopping parameter is thus given by $\Kc = 1/(8+2a\mc)$ .

With these definitions 
the renormalized improved bilinear operators  
take the form \cite{paperI} (${\rm X=V,A,P,T}$)
\begin{equation}
 X_{\rm R}=Z_{\rm X}(\tilde{g}_0^2,a\mu)
           \bigl[1+b_{\rm X}(g_0^2) a\mq\bigr]X_{\rm I}.
\end{equation}

For the Wilson plaquette action a non-perturbative determination
in the quenched approximation  
of $\Za, \bv$ can be found in \cite{paperIV}; of $\ca$ in 
\cite{ca}; and of $\cv$ in \cite{MarcoRainer}.

\subsection{Computation of the improvement coefficients}

One first considers the massless theory. In this case
on-shell correlation functions involving only 
the composite fields are O($a$) improved provided 
the improvement coefficients $\csw$, $\ca$, $\cv$ and $\cT$
are properly chosen as functions of the bare coupling $g_0$. 
In the Schr\"{o}dinger functional framework it is, however, 
usually easier 
to perturbatively compute correlation functions also
involving the near-boundary quark fields $\zeta,\zetabar$ 
and $\zeta^{\prime},\zetabar^{\prime}$~ \cite{paperI,paperII}. 
O($a$) improved correlation functions
are then obtained with the renormalized fields 
$\zeta_{\rm R},\zetabar_{\rm R}$ and
$\zeta_{\rm R}^{\prime},\zetabar_{\rm R}^{\prime}$,
which are all related to the bare fields
by the  same renormalization factor, e.g.
\begin{equation}
  \zeta_{\rm R}=\zzeta(\tilde{g}_0^2,a\mu)
	 	  \bigl[1+\bzeta(g_0^2) a\mq\bigr]\zeta. 
\end{equation}

The particular set of correlation functions which we have chosen
to compute was precisely as for the case of the Wilson action
described in detail in refs.\cite{paperII, StefanPeter}. Our
calculation determines of course only the 1-loop correction to
the tree level values of the improvement coefficients, i.e. the
quantities $c_{\rm X}^{(1)}$ in the perturbative expansion for a
general improvement coefficient $c_{\rm X}$:
\begin{equation} \label{c_X_expansion}
c_{\rm X} = c_{\rm X}^{(0)} + c_{\rm X}^{(1)} g_0^2 
+ c_{\rm X}^{(2)}g_0^4 + \dots
\end{equation}

Only for the computation of
$\csw^{(1)}$ is it necessary to have a non-vanishing background
field. Once one knows this coefficient the others can be determined
by computing correlation functions in the absence of a background 
field, for which the programs are considerably less CPU time consuming.
In the case of vanishing quark mass ($m_{\rm R}L$=0) we have collected
data at $T=2L$, for $\theta=0$ and $\theta=1$, with vanishing background
field and at $T=L$, only for $\theta=0$, with non-vanishing background field.

After calculating the improvement coefficients which are
necessary in the massless theory one can proceed to
the determination of the 1-loop contribution to the various
$b$-coefficients that were introduced above and are required for 
the theory with massive quarks. These coefficients have an expansion
in powers of $g_0^2$ of the same form as the expansion
(\ref{c_X_expansion}) for the c-coefficients.
In the massive case one has to study quark correlation functions at fixed
values of $\mr L$. The set of correlation functions that we considered
is exactly the same as described in detail in ref.~\cite{StefanPeter}.
Also in this case the computation can be done at vanishing gauge 
background field. Data were collected for two values of $m_{\rm R}L$,
namely $0.1$ and $0.5$ and $\theta =0$, by taking $T=2L$. 

We remark that we did not directly determine the coefficient $\bs^{(1)}$
for the isovector scalar operator $S^a$ since it can be 
determined by the relation 
\begin{equation} 
\bs^{(1)}= -2 \bm^{(1)} .
\end{equation}
In the quenched
approximation one can prove the non-perturbative relation $\bs=-2 \bm$
\footnote{
A brief outline of the proof, which is due to M. L\"{u}scher,
can be found in the last section of ref.\cite{Weisz}.
Please note that in this paper the
factor on the rhs of eq.(113) should be $(1+2\bm a\mq)^{-1}$.}
from which the 1-loop relation above follows for the fully interacting
theory. We also observe that the improvement coefficient $\bg$
(see Eq.(2.44)) vanishes at tree level and at one loop
order it is independent of the pure gauge action:
$\bg^{(1)}=0.012000(2)\Nf$ \cite{StefanRainer}.

It should also be noticed that the values of $\ctildet^{(1)}$ 
and $\bzeta^{(1)}$ actually needed
for the O($a$) improvement of correlation functions involving boundary
quark fields depends in general on the precise form 
of the pure gauge action. This involves not 
only the different choices of the Wilson loop coefficients $c_i$
($i=0,1,2,3$),
but also the particular choice made for the SF setup. 
Our results for $\ctildet^{(1)}$
and $\bzeta^{(1)}$ refer specifically to 
the above mentioned Choice A for 
all the considered pure gauge actions.


\section{Analysis Method}
We describe in this section the method that we adopted to extrapolate
the asymptotic values of the various improvement coefficients from the
1-loop results obtained on a sequence of finite lattices.

The 1-loop results were produced by Fortran codes in double precision.
We have checked the gauge invariance of the results produced with our
codes by adopting different covariant
gauge fixings within the family of eq.(\ref{gaugefix}). Checks were also
done against pre-existing safe codes in some limiting cases, like the
case of the standard Wilson plaquette action or the case of vanishing
background field. Further checks about the correctness of the codes
and of the whole theoretical setup were obtained {\em a posteriori}
by the observed consistency of independent determinations of the same 
improvement coefficients.

The effects of rounding errors and their dependence on the lattice size
have been investigated by comparing, for a limited set of cases, results
obtained by running the same code in single, double and extended precision.
From this investigation we have derived a formula which gives a rough
estimate of the rounding-error effects as a function of the lattice size.
The estimated rounding errors were conservatively propagated through the
whole analysis procedure described below. In all the cases, when the analysis
output was clearly affected by rounding errors, our estimate was
found to be realistic or even conservative. 

The starting point of the analysis procedure are the (gauge invariant)
correlation functions $f_{\rm X}(L;T/L;\theta;m_{\rm R}L)$, evaluated
on a sequence of lattices of volume $L^3 \times T$ at
a given fixed value of $m_{\rm R}L$ and with the spatial boundary conditions
for the fermion field specified by $\theta$, as explained in section 2.
Here X labels the different quark correlation functions within the
set 
considered in refs. \cite{paperII, StefanPeter}.

Taking suitable linear combinations of these correlation functions,
possibly involving different values of $\theta$ and $m_{\rm R}L$, 
enables us
to construct lattice approximants of the improvement coefficients or some
simple linear combinations of them, which are the quantities that
we will consider from now on. The precise form of these quantities is
not given here, because, as already mentioned, they are exactly the same
as specified in \cite{paperII}, for the determination of $\Kc^{(1)}$,
$\ca^{(1)}$, $\ctildet^{(1)}$ and $\csw^{(1)}$, and in
\cite{StefanPeter} 
for all the remaining improvement coefficients introduced above.

Each sequence $\{ Q \}$ of lattice approximants has, by construction,
a well definite limit as $L \to \infty$ and a large $L$ expansion of the
form:
\begin{equation} \label{large_L_form}
Q(L;T/L) = Q_0 + \sum_{i=1}^{\infty} [Q_i + q_i \log(L)] L^{-i} \; ,
\end{equation}
where the lattice artifacts coefficients $Q_i$ and $q_i$ depend on
the value of $T/L$. We are of course interested in the asymptotic
value ($Q_0$) of the sequence, which is obtained by suitable extrapolation
to the limit $L \to \infty$, performed at a fixed the value of $T/L$.

In order to get the asymptotic value of the various sequences of 
interest we have used the method of \cite{LueWe}, suitably adapted
to the case of functions of $L$ of the form (\ref{large_L_form}). 
The main
point of the method is to perform some consecutive blocking steps
on the initial data, represented by the sequence $\{ Q \}$. In the first
step, for each value of $L$, one constructs the linear combination of the data 
at $L$ and the two contiguous values 
of the lattice size ($L-1$ and $L+1$, in the case $T=2L$)
that does not contain terms of order $L^{-1}$ and $\log(L) L^{-1}$. 
In the second step, the terms of order $L^{-2}$ and $\log(L) L^{-2}$ are
analogously removed and so on. At each step the number of data in the
resulting sequence is of course decreased by two units and the 
rounding-error effects are enhanced, which limits the optimal number
of steps to be performed in practice. This method allows for a safe, 
although often conservative, estimate of the systematic error on 
the asymptotic value of the sequence under study. To this aim it
is important to properly propagate the estimated rounding errors on
$f_{\rm X}(L;T/L;\theta;\mr L)$ when constructing all the needed
linear combinations. 
Our practical use of the method closely followed the
original proposal and detailed description in \cite{LueWe}, apart 
from some details which are discussed below. 

The data for the correlation functions 
$f_{\rm X}(L;T/L;\theta;\mr L)$ 
were collected for values of $L$ ranging from $4$ to $L_{max}$, as indicated
in table \ref{tab:result}. 
Because our data have been obtained using only double precision and with
a value of $L_{max}=32$, the uncertainties on the values of improvement
coefficients that we quote for LW, RG1, RG2 and RG3 actions are systematically 
larger than the uncertainties obtained in \cite{paperII, StefanPeter}
for the standard plaquette action. In addition, we had also to 
adopt a slightly modified blocking--fit procedure, in some cases
where a straightforward application of the above mentioned blocking procedure
turned out to lead to a large (order $10\%$) estimated
value of the relative  
error on the asymptotic quantities we are interested in.  

We realized, indeed, that when the lattice artifacts of a given order $j$
in the asymptotic expansion (\ref{large_L_form}), i.e.
\[ [Q_j + q_j \log(L)] L^{-j} \] 
are numerically very small 
(in comparison with higher order lattice artifacts)
for all the considered values of $L$, 
removing them through a suitable linear combination
strongly enhances rounding-error effects while hardly changing the
estimate of the asymptotic value ($Q_0$ in (\ref{large_L_form})). In such
a case we found it convenient for the purpose of improving the precision
on $Q_0$, to resort to the following hybrid blocking--fit procedure:
\begin{itemize}
\item  perform standard blocking steps for terms of order from $0$ to $j-1$;
\item  using suitable linear combinations of the reduced data sequence at
       step $j-1$, determine separately the values of $Q_j$ and $q_j$, 
       say $\bar{Q}_j$ and $\bar{q}_j$, with some (in practice
       rather large) relative uncertainty $\delta Q_j$ and $\delta q_j$;
\item  subtract from the reduced data sequence at step $j-1$ the quantity
       $[\bar{Q}_j + \bar{q}_j \log(L)] L^{-j}$ and perform standard
       blocking steps for terms of order $j+1$, $j+2$, \dots, stopping
       as usual when the number of data in the sequence gets too small
       or the rounding-error effects too large
\item  repeat the procedure of the latter item
       by replacing $\bar{Q}_j$ and $\bar{q}_j$ with
       $\bar{Q}_j \pm \delta Q_j$ and $\bar{q}_j \pm \delta q_j$, respectively,
       taking mean and extreme values in a way that is equivalent to the error 
       estimate adopted for standard blocking steps. In this way one gets
       a conservative estimate of the uncertainty induced by $\delta Q_j$
       and $\delta q_j$. 
\end{itemize}

This blocking--fit procedure is equivalent to the simple blocking 
procedure of \cite{LueWe}
in the ideal case of infinite precision data, 
but turns out to be more convenient
when rounding-error effects are not negligible and lattice artifacts of a 
given order $j$ are very small.

We remark that requiring O($a$) improvement of all the correlation
functions
introduced in ref.s \cite{paperII, StefanPeter} provides a number of
relations between the various improvement coefficients larger than the number
of the improvement coefficients themselves, allowing then for several 
consistency checks. Further consistency checks were done by comparing the
values of the b-coefficients at $\mr L=0.1$ and $\mr L=0.5$.

After checking the consistency among the asymptotic values obtained from 
different lattice approximants of the same improvement coefficients, we
have sometimes made use of the fact that a particular 
linear combination of these 
approximants showed small lattice artifacts, leading to a more precise
determination of the asymptotic value than the single approximants themselves.

\section{Results and Conclusion}
In table \ref{tab:result} we give our final result obtained
by the method in the previous section. 
This table contains a synthesis of all our 1-loop 
results, including a comparison with the case of the standard plaquette gauge
action (label ``Plaq.'')\cite{paperII,StefanPeter}. 

We first notice that tadpole contributions to the
$\csw^{(1)}$, denoted $\csw^{tad}$ in the table, give about 90\% of 
the complete 1-loop contributions for all actions considered here.
Therefore the value of $\csw$ taken by the CP-PACS collaboration for their
full QCD simulation with RG1 gauge action\cite{CppacsF} is very close to 
the full one-loop value up to order $g_0^4$, and the order $g_0^2$
difference can be calculated from the table: 
$\csw^{pert.} = \csw^{CP-PACS}+0.008 g_0^2 +{\rm O}(g_0^4)$ .

Although only 3 choices of RG improved gauge actions are considered,
it seems that RG improved gauge actions generally give 
a $\csw^{(1)}$ coefficient which is a factor 2 to 2.5 smaller
than that of the plaquette action,
while for the perturbative improved action (LW)
the reduction factor is only about $1.35$.
This tendency has already been found in the finite 
part of 1-loop renormalization factors for various quantities \cite{ATNU}.
Other improvement coefficients, the $c_{\rm X}^{(1)}$'s
and $b_{\rm X}^{(1)}$'s, as well as
$\bzeta^{(1)}$ and $\ctildet^{(1)}$ 
for the setup denoted as Choice A, are 
also given in table \ref{tab:result}.
The same tendency for the size of 1-loop coefficients is also observed.

\begin{figure}[htb]
\begin{center}
\leavevmode
\epsfxsize=90mm
\epsfbox{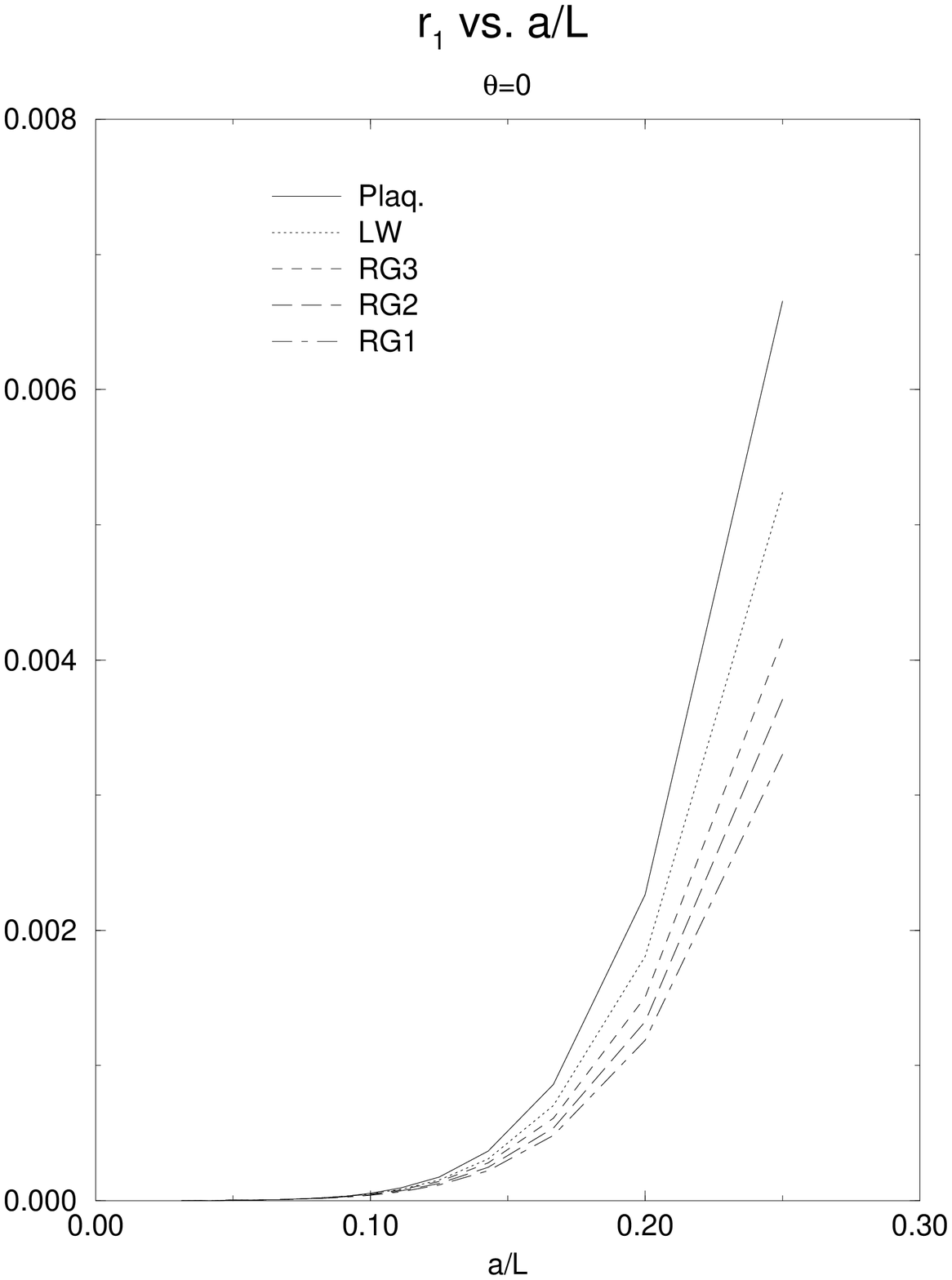}
\vskip 10mm
\end{center}
\caption{Remaining cutoff effect at 1-loop in the PCAC mass at
$x_0=T/2$ and $K = \Kc$ for various 
actions, for the case $T=2 L$, $\theta =0$ and vanishing background
field.}
\label{fig:pcac}  
\end{figure}  

It is clear, however, that the smallness of 1-loop coefficient for the
RG improved action does not imply the smallness of the lattice artifacts
for the action. 
To get some impression of these we have to evaluate
lattice artifacts directly from physical observables for the corresponding
action. For that purpose, following ref.~\cite{paperII}, we consider the 
unrenormalized current quark mass $m$, defined through the PCAC relation
(see eq.(6.13) of ref.~\cite{paperI} ), which is expected to vanish up to 
terms of order $a^2$ at $K = \Kc$.
If we expand $m$ such that
\begin{equation}
 am  = r_0 + r_1 \cf g_0^2 + {\rm O}(g_0^4) 
\end{equation}
at $K=\Kc$ and $x_0=T/2$, 
the value of $r_1$ represents the magnitude of the remaining
cutoff effect for the particular gauge action at 1-loop order,
since the tree-level contribution 
$r_0$ is independent of the pure gauge action.
In fig.~\ref{fig:pcac} we have plotted $r_1$ for various actions
as a function of $a/L$ for $T= 2 L$ and $\theta = 0$ and 
vanishing boundary gauge field (in which case 
the tree-level contribution actually vanishes). 
At large values of $a/L$ it is observed that this
1-loop lattice artifact 
is indeed smaller for the RG improved gauge actions than
for the LW action, for which it is still smaller than 
for the plaquette action.
We remark however that one cannot draw general conclusions
from this simple exercise; the pattern may be different for other 
correlation functions and, needless to say, at a non-perturbative 
level we have no statement to make at all.

In the last box
of the table the 1-loop value of the critical hopping parameter
$\Kc^{(1)}$, defined by $\Kc = \Kc^{(0)} + g_0^2 \Kc^{(1)}+ \cdots$, is
also given and compared with the results from perturbation
theory on an infinite lattice \cite{ATNU}. 
They agree with each other within 5 digits, and this fact also
supports the correctness of our calculations.

Recently the coefficients $c_{\rm X}$ and $b_{\rm X}$ have been 
calculated
by Taniguchi and Ukawa \cite{TaUk} using a completely different method. 
The two sets of results agree well, 
although our quoted errors are generally larger than theirs.
The good agreement between the central values 
of the two methods indicates, however,
that our error estimates are rather conservative.

\section*{Acknowledgements}

The authors would like to thank Martin L\"{u}scher 
for crucial discussions.
S.A. thanks Y. Taniguchi and A. Ukawa for informative correspondence.


\renewcommand{\theequation}{\mbox{A.\arabic{equation}}}
\section*{Appendix A}

In this appendix we explain the allowable weights of the
contributions to the gauge action near the boundary in our 
setup to avoid order $a$ effects. We remind the reader that in this 
paper we are not demanding the absence of O($a^2$) effects.
As explained in ref.~\cite{alphaI} there are only two
possible boundary counterterms 
at order $a$ in the pure gauge theory, which are obtained by 
summing any local lattice expressions for the fields
\begin{equation}
a^4\tr \Bigl\{F_{0k}F_{0k}\Bigr\}\,\,\, {\rm and}\,\,\,
a^4\tr \Bigl\{F_{ij}F_{ij}\Bigr\}
\end{equation} 
over the $x_0=0$ and $x_0=T$ hyperplanes. 
 
The first step is to ensure that the approach to
the classical continuum limit is O($a^2$).
Following the discussion of  ref.~\cite{LueWeI} 
we recall that any given smooth continuum gauge
field $A_{\mu}(y)$, can be arbitrarily well approximated
by lattice gauge fields $U_{\mu}(x)$
\begin{equation}
U_{\mu}(x)=T\exp \Bigl\{a\int_0^1 \rmd t A_{\mu}(x+[1-t]a\hat{\mu})
\Bigr\}\,.
\end{equation}
One can then derive asymptotic expansions for the lattice operators.
For example for 
${\cal L}({\cal C}_i)={\rm Re Tr}[I-U({\cal C}_i)]$ with 
$C_i\in {\cal S}_i$  
one  easily finds for the flat loops in the $\mu-\nu$ plane
\begin{equation}\label{flat}
{\cal L}({\cal C}_i)=-k_i a^4\tr F_{\mu\nu}(x_c)F_{\mu\nu}(x_c)
+{\rm O}(a^6),\,\, i=0,1
\end{equation}
where the constant $k_i$ depends on the class
\begin{equation}
k_0=\frac12,\,\,\,,k_1=2\,.
\end{equation}
and $x_c$ is the ``center of the loop" (otherwise one obtains 
in general O($a^5$) appearing on the rhs of Eq.(\ref{flat}) above).
Obviously there are similar expressions for the non-planar loops. 

The approximation of the integrals by sums in spatial directions 
can incur only O($a^2$)
errors because of the periodic boundary conditions. 
However approximations of the integral by sums in the
time direction may 
result in additional effects of order $a$ through boundary terms. 
For example for a smooth function $f(z)$ defined on the
region $[0,T]$ and dividing the region in $M=T/a$ segments, we
have
\begin{eqnarray}\label{sums}
a\sum_{m=1}^M f(ma-\frac12 a)&=& \int_0^T \rmd z f(z) + {\rm O}(a^2),
\\
\noalign{\vskip2ex}\label{sums2}
a\sum_{m=1}^{M-1} f(ma) &=& \int_0^T \rmd z f(z) 
-{a\over2}\Bigl(f(0)+f(T)\Bigr) + {\rm O}(a^2).
\end{eqnarray}

For planar loops extending in the time direction for which the
time coordinate of the center $x_c$ takes half 
integer values we can invoke Eq.(\ref{sums}) to see that their
contribution to the action will be as in the continuum
up to terms of order $a^2$.
However for those loops where $x_c$ takes integer values,
which is the case only for  
the rectangles extending over two time units, we see from (\ref{sums2})
that the contribution to the action from these rectangles will have
an O($a$) deficit with respect to the classical continuum expression.
For instance at the boundary $T=0$ we are missing a contribution
$-(2c_1/g_0^2)\int \rmd {\bf x} \tr F_{0k}(0,{\bfx})F_{0k}(0,{\bfx})$ 
which can be compensated by adding extra terms involving plaquettes 
and/or rectangles extending only one unit in the time direction.
Two admissible choices are the special setups A, B specified in
subsection 2.2.

Eq.(\ref{sums2}) also explains the fact that $\cs(0)=\frac12$, but this is 
not relevant for our particular boundary conditions for which
$F_{jk}=0$.

\renewcommand{\theequation}{\mbox{B.\arabic{equation}}}
\section*{Appendix B}

In this appendix we give the form of the free inverse gauge propagator
for non-zero $c_1$ and $c_3$ in the case of vanishing background field.
Introducing the Fourier transformed fields through
\begin{eqnarray}
q_0(x)&=&L^{-3}\sum_{\bfp}e^{i\bfp\bfx}\tilde{q}_0(\bfp,x_0), \\
q_k(x)&=&L^{-3}\sum_{\bfp}e^{i\bfp\bfx}e^{ip_k/2}\tilde{q}_k(\bfp,x_0) ,
\end{eqnarray}
the quadratic part of the gauge
action including the gauge-fixing term takes the form
\begin{equation}
S_{\rm gauge}^{(2)} = -L^{-3}\sum_\bfp\sum_{x_0,y_0=0}^{T-1}
\tr \tilde{q}_\mu(-\bfp,x_0)\tilde{q}_\nu(\bfp,y_0)
K_{\mu\nu}(\bfp;x_0,y_0) .
\end{equation}
Explicitly
\begin{eqnarray}
K_{00}(\bfp;x_0,y_0) &=&
\delta_{x_0,y_0}\Bigl[
(c_0+6c_1+8c_3)\hat\bfp^2-c_1\hat\bfp^4+c_3(\hat\bfp^4-(\hat\bfp^2)^2) 
\nonumber\\
&+ &
\lambda_0\{ 2-\delta_{x_0,0}(1-\delta_{\bfp,\bf 0})-\delta_{x_0,T-1}\}
\Bigr] \nonumber\\
&+& (\delta_{x_0-1,y_0}+\delta_{x_0+1,y_0})(c_1\hat\bfp^2-\lambda_0)
\nonumber
\\
&+&c_1\delta_{x_0,y_0}(\delta_{x_0,T-1}+\delta_{x_0,0})(\hat\bfp^2-\hat\bfp^4
d_{bc}) , \\
K_{k0}(\bfp;x_0,y_0) &=&
i\hat{p_k}\Bigl[(\delta_{x_0,y_0}-\delta_{x_0-1,y_0})
\{c_0+5c_1+8c_3-\lambda_0-c_1\hat{p_k}^2+c_3(\hat{p_k}^2-\hat\bfp^2)\}
\nonumber\\
&+& 
c_1(\delta_{x_0+1,y_0}-\delta_{x_0-2,y_0}) \nonumber \\
&+& c_1 (\delta_{x_0,y_0}\delta_{x_0,T-1}-\delta_{x_0-1,y_0}\delta_{x_0,1})
(1-\hat{p_k}^2 d_{bc})
\Bigr] , \\
K_{0k}(\bfp;x_0,y_0) &=&-K_{k0}(\bfp;y_0,x_0), \\
K_{kl}(\bfp;x_0,y_0) &=& \delta_{x_0,y_0}\delta_{kl}\Bigl[
(c_0+8c_1+4c_3)\hat\bfp^2 + 2c_0+10c_1+16c_3
\nonumber\\ 
&-&
c_1(2\hat{p_k}^2+\hat\bfp^4+\hat\bfp^2\hat{p_k}^2) 
+c_3(2\hat{p_k}^2+\hat\bfp^4-(\hat\bfp^2)^2+\hat\bfp^2\hat{p_k}^2)
\Bigr] \nonumber \\
&+& \delta_{x_0,y_0}\hat{p_k}\hat{p_l}\Bigl[
\lambda_0-c_0-8c_1-6c_3+c_1(\hat{p_k}^2+\hat{p_l}^2)+
c_3(\hat\bfp^2-\hat{p_k}^2-\hat{p_l}^2) 
\Bigr]\nonumber \\
&-&(\delta_{x_0+1,y_0}+\delta_{x_0-1,y_0})\Bigl[
\delta_{kl}\{c_0+4c_1+8c_3-c_1\hat{p_k}^2+c_3(\hat{p_k}^2-2\hat\bfp^2)\}
\nonumber \\
&+&
c_3\hat{p_k}\hat{p_l}
\Bigr] 
-(\delta_{x_0+2,y_0}+\delta_{x_0-2,y_0})\delta_{kl}c_1 \nonumber\\
&+& c_1\delta_{kl}\delta_{x_0,y_0}(\delta_{x_0,T-1}+\delta_{x_0,1})
(1-\hat{p_k}^2 d_{bc}) ,
\end{eqnarray}
where
\begin{equation}
\hat{p_k} =2 \sin p_k/2 , \qquad
\hat\bfp^2 =\sum_{k=1}^3 \hat{p_k}^2 , \qquad
\hat\bfp^4 =\sum_{k=1}^3 \hat{p_k}^4 .
\end{equation}
For the boundary coefficient $d_{bc}$ appearing above, we have
$d_{bc}=0$ for Choice A and  $d_{bc}=1$ for Choice B.


\vfill
\eject
\section*{Table}
\begin{table}[bht]
\caption{Improvement coefficients at 1-loop for various gluon actions.
The results for the standard plaquette action are taken from
refs.{\protect
\cite{paperII,StefanPeter}}. All results are obtained from data with $T = 2 L$
except $\csw^{(1)}$, which is calculated from $T=L$ data.
Tadpole contributions for $\csw^{(1)}$ are also listed. 
The critical hopping parameter $\Kc^{(1)}$ (above) is also compared with 
the result (below) obtained from the usual perturbation theory in 
infinite volume {\protect\cite{ATNU}}.
}
\label{tab:result}
\begin{center}
\begin{tabular}{|c|lllll|}
\hline
gauge action& Plaq. & LW & RG1 & RG2 & RG3  \\
\hline
$c_1$ &0.0 &-1/12 &-0.331 & -0.27 &-0.252 \\
$c_3$ &0.0 &\phantom{-}0.0    &\phantom{-}0.0   & -0.04 &-0.17  \\
\hline
&\multicolumn{5}{c|}{} \\
\hline
$L_{max}$      & 32       & 24    & 30    & 24    & 24    \\
$\csw^{(1)}$       & 0.267(1) & 0.196(6) & 0.113(3) & 0.119(5) & 0.109(5)
\\
$\csw^{tad}$ & 0.25     & 0.183 & 0.105 & 0.110 & 0.096 \\
\hline
$L_{max}$      & 48       & 32    & 32    & 32    & 32    \\
&\multicolumn{5}{c|}{$\times \cf$}   \\
$\ca^{(1)}$ &-0.005680(2)&-0.004525(25) &-0.002846(11) &-0.003017(12) &
-0.002805(20)\\
$\cv^{(1)}$ &-0.01225(1)
&-0.0103(3)&-0.00730(20)&-0.00757(26)&-0.00709(20)\\
$\cT^{(1)}$ &\phantom{-}0.00896(1) &\phantom{-}0.00743(7)
&\phantom{-}0.00505(10) &\phantom{-}0.00526(15) &\phantom{-}0.00496(12) \\
$\bm^{(1)}$ &-0.07217(2)&-0.0576(11)&-0.0382(8)&-0.0395(15) &-0.0353(12)
\\
$\ba^{(1)}$
&\phantom{-}0.11414(4)&\phantom{-}0.0881(13)&\phantom{-}0.0550(4)
&\phantom{-}0.0572(6) &\phantom{-}0.0510(5)  \\
$\bv^{(1)}$
&\phantom{-}0.11492(4)&\phantom{-}0.0884(26)&\phantom{-}0.0551(19)
&\phantom{-}0.0574(19)&\phantom{-}0.0510(21) \\ 
$\bp^{(1)}$
&\phantom{-}0.11484(2)&\phantom{-}0.0889(14)&\phantom{-}0.0558(9)
&\phantom{-}0.0584(10)&\phantom{-}0.0528(8) \\
$\bs^{(1)}$
&\phantom{-}0.14434(4)&\phantom{-}0.1152(22)&\phantom{-}0.0764(16)
&\phantom{-}0.0790(30)&\phantom{-}0.0706(24)\\
$\bt^{(1)}$
&\phantom{-}0.10434(4)&\phantom{-}0.0790(25)&\phantom{-}0.0477(12)
&\phantom{-}0.0502(19)&\phantom{-}0.0444(15) \\ 
$\bzeta^{(1)}$
&-0.06738(4)&-0.0505(23)&-0.0243(11)&-0.0275(18)&-0.0252(11) \\
$\ctildet^{(1)}$ &-0.01346(1)&-0.0122(4)&-0.00661(21)&-0.00612(23)
&-0.00186(20) \\
\hline
$L_{max}$      & 32       & 32    & 32    & 32    & 32    \\
& \multicolumn{5}{c|}{$\times \cf\times 10^{-3}$} \\
$\Kc^{(1)}$& 6.329891(3) &4.716253(1) &2.760894(1)&2.911363(2)&2.593490(1)\\
$\Kc^{(1)}$  & 6.3300(1)  &4.7163(1)  & 2.7609(1)  & 2.9114(1) & 2.5935(1) \\
\hline
\end{tabular}
\end{center}
\end{table}

\end{document}